\documentclass[dvips]{article}
\usepackage{supertabular,lscape,epsfig}
\usepackage{amssymb}
\usepackage{amsmath}

\DeclareSymbolFont{ppa}{OT1}{ppl}{m}{it}
\DeclareMathSymbol{\vv}{\mathalpha}{ppa}{'166}

\begin{document}

\newcommand{\dd}{\,{\rm d}}
\newcommand{\ie}{{\it i.e.},\,}
\newcommand{\etal}{{\it et al.\ }}
\newcommand{\eg}{{\it e.g.},\,}
\newcommand{\cf}{{\it cf.\ }}
\newcommand{\vs}{{\it vs.\ }}
\newcommand{\zdot}{\makebox[0pt][l]{.}}
\newcommand{\up}[1]{\ifmmode^{\rm #1}\else$^{\rm #1}$\fi}
\newcommand{\dn}[1]{\ifmmode_{\rm #1}\else$_{\rm #1}$\fi}
\newcommand{\upd}{\up{d}}
\newcommand{\uph}{\up{h}}
\newcommand{\upm}{\up{m}}
\newcommand{\ups}{\up{s}}
\newcommand{\arcd}{\ifmmode^{\circ}\else$^{\circ}$\fi}
\newcommand{\arcm}{\ifmmode{'}\else$'$\fi}
\newcommand{\arcs}{\ifmmode{''}\else$''$\fi}
\newcommand{\MS}{{\rm M}\ifmmode_{\odot}\else$_{\odot}$\fi}
\newcommand{\RS}{{\rm R}\ifmmode_{\odot}\else$_{\odot}$\fi}
\newcommand{\LS}{{\rm L}\ifmmode_{\odot}\else$_{\odot}$\fi}

\newcommand{\Abstract}[2]{{\footnotesize\begin{center}ABSTRACT\end{center}

\vspace{1mm}\par#1\par
\noindent
{~}{\it #2}}}

\newcommand{\TabCap}[2]{\begin{center}\parbox[t]{#1}{\begin{center}
  \small {\spaceskip 2pt plus 1pt minus 1pt T a b l e}
  \refstepcounter{table}\thetable \\[2mm]
  \footnotesize #2 \end{center}}\end{center}}

\newcommand{\TableSep}[2]{\begin{table}[p]\vspace{#1}
\TabCap{#2}\end{table}}

\newcommand{\FigCap}[1]{\footnotesize\par\noindent Fig.\  %
  \refstepcounter{figure}\thefigure. #1\par}

\newcommand{\TableFont}{\footnotesize}
\newcommand{\TableFontIt}{\ttit}
\newcommand{\SetTableFont}[1]{\renewcommand{\TableFont}{#1}}

\newcommand{\MakeTable}[4]{\begin{table}[htb]\TabCap{#2}{#3}
  \begin{center} \TableFont \begin{tabular}{#1} #4 
  \end{tabular}\end{center}\end{table}}

\newcommand{\MakeTableSep}[4]{\begin{table}[p]\TabCap{#2}{#3}
  \begin{center} \TableFont \begin{tabular}{#1} #4 
  \end{tabular}\end{center}\end{table}}

\newcommand{\TabCapp}[2]{\begin{center}\parbox[t]{#1}{\centerline{
  \small {\spaceskip 2pt plus 1pt minus 1pt T a b l e}
  \refstepcounter{table}\thetable}
  \vskip0mm
  \centerline{\footnotesize #2}}
  \vskip0mm
\end{center}}

\newcommand{\MakeTableSepp}[4]{\begin{table}[p]\TabCapp{#2}{#3}
  \begin{center} \TableFont \begin{tabular}{#1} #4
  \end{tabular}\end{center}\end{table}}

\newenvironment{references}%
{
\footnotesize \frenchspacing
\renewcommand{\thesection}{}
\renewcommand{\in}{{\rm in }}
\renewcommand{\AA}{Astron.\ Astrophys.}
\newcommand{\AAS}{Astron.~Astrophys.~Suppl.~Ser.}
\newcommand{\ApJ}{Astrophys.\ J.}
\newcommand{\ApJS}{Astrophys.\ J.~Suppl.~Ser.}
\newcommand{\ApJL}{Astrophys.\ J.~Letters}
\newcommand{\AJ}{Astron.\ J.}
\newcommand{\IBVS}{IBVS}
\newcommand{\PASP}{P.A.S.P.}
\newcommand{\Acta}{Acta Astron.}
\newcommand{\MNRAS}{MNRAS}
\renewcommand{\and}{{\rm and }}
\section{{\rm REFERENCES}}
\sloppy \hyphenpenalty10000
\begin{list}{}{\leftmargin1cm\listparindent-1cm
\itemindent\listparindent\parsep0pt\itemsep0pt}}%
{\end{list}\vspace{2mm}}

\def\TYLDA{~}
\newlength{\DW}
\settowidth{\DW}{0}
\newcommand{\dw}{\hspace{\DW}}

\newcommand{\refitem}[5]{\item[]{#1} #2%
\def\REFARG{#3}\ifx\REFARG\TYLDA\else, {\it#3}\fi
\def\REFARG{#4}\ifx\REFARG\TYLDA\else, {\bf#4}\fi
\def\REFARG{#5}\ifx\REFARG\TYLDA\else, {#5}\fi.}

\newcommand{\Section}[1]{\section{\hskip-6mm.\hskip3mm#1}}
\newcommand{\Subsection}[1]{\subsection{#1}}
\newcommand{\Acknow}[1]{\par\vspace{5mm}{\bf Acknowledgements.} #1}
\pagestyle{myheadings}

\newfont{\bb}{ptmbi8t at 12pt}
\newcommand{\xrule}{\rule{0pt}{2.5ex}}
\newcommand{\xxrule}{\rule[-1.8ex]{0pt}{4.5ex}}
\def\thefootnote{\fnsymbol{footnote}}

\begin{center}
{\Large\bf The Optical Gravitational Lensing Experiment.
\vskip3pt
The New Catalog of Eclipsing Binary Stars 
\vskip5pt
in the Small Magellanic Cloud.\footnote{Based on observations
obtained with the 1.3~m Warsaw telescope at the Las Campanas  
Observatory of the Carnegie Institution of Washington.}}
\vskip 3mm
{\bf {\L}.~~W~y~r~z~y~k~o~w~s~k~i$^{1,2}$\!,
~~A.~~U~d~a~l~s~k~i$^1$\!, ~~M. ~~K~u~b~i~a~k$^1$\!,
~~M.\,K.~~S~z~y~m~a~{\'n}~s~k~i$^1$\!,~~ K.~~{\.Z}~e~b~r~u~{\'n}$^1$\!,
~~I.~~S~o~s~z~y~{\'n}~s~k~i$^1$\!,\\ P.\,R.~~W~o~{\'z}~n~i~a~k$^3$\!,
~~G.~~P~i~e~t~r~z~y~{\'n}~s~k~i$^{1,4}$~~and 
~~O.~~S~z~e~w~c~z~y~k$^1$}
\vskip 2mm
{$^1$ Warsaw University Observatory, Al.~Ujazdowskie~4, 00-478~Warsaw, Poland\\
e-mail: (wyrzykow,udalski,mk,msz,zebrun,soszynsk,pietrzyn,szewczyk)@astrouw.edu.pl\\
$^2$ School of Physics and Astronomy and Wise Observatory, Tel-Aviv University, Tel~Aviv~69978, Israel\\
e-mail: lukas@wise.tau.ac.il\\
$^3$ Los Alamos National Observatory, MS-D436, Los Alamos NM 85745, USA\\
e-mail: wozniak@lanl.gov\\
$^4$ Universidad de Concepci{\'o}n, Departamento de Fisica,
Casilla 160-C, Concepci{\'o}n, Chile\\
e-mail: pietrzyn@hubble.cfm.udec.cl}
\end{center}

\vspace*{9pt}
\Abstract{We present new version of the OGLE-II catalog of eclipsing
binary stars detected in the Small Magellanic Cloud, based on Difference
Image Analysis catalog of variable stars in the Magellanic Clouds
containing data collected from 1997 to 2000.

We found 1351 eclipsing binary stars in the central 2.4 square degree
area of the SMC. 455 stars are newly discovered objects, not found in
the previous release of the catalog. The eclipsing objects were selected
with the automatic search algorithm based on the artificial neural
network. The full catalog is accessible from the OGLE {\sc Internet}
archive.}{Keywords: binaries: eclipsing -- Magellanic Clouds -- Catalogs} 

\Section{Introduction}
Precise determination of distances to nearby galaxies is still one of
the main goals of modern astrophysics. Eclipsing binary stars were used
for this purpose for almost hundred years and in the last decade we
witnessed their great ``comeback'' for two main reasons. First, very
large telescopes with mirror diameter more than 6 meters can provide
accurate spectroscopy of such faint stars as eclipsing binaries in
nearby galaxies. Secondly, long time-base photometry and precise light
curves of eclipsing binaries, mostly in Magellanic Clouds and Galactic
bulge, are supplied as a by product of microlensing searches, \eg MACHO
(Alcock \etal 1997) and OGLE (Udalski \etal 1997a, Udalski \etal 1998,
Wyrzykowski \etal 2003).

To date several attempts of distance determination with eclipsing binary
method were presented (\eg Fitzpatrick \etal 2003, Fitzpatrick \etal
2002, Ribas \etal 2002), mostly to the LMC because this value is crucial
for the distance scale. The extragalactic distance scale is tied to the
LMC distance.

In last few years, the distance to the Small Magellanic Cloud was also
determined with eclipsing binary method by several authors: Wyithe and
Wilson (2001), Wyithe and Wilson (2002), Harries, Hilditch and Howarth
(2003). Their papers were based on the photometry obtained by the OGLE
collaboration (Udalski \etal 1998), which contained data from the first
1.5 year of observations of the second phase of the OGLE survey
(Udalski, Kubiak and Szyma{\'n}ski 1997b). However, OGLE-II continued
collecting data until the end of 2000. A much larger and almost complete
subset of the OGLE-II images was reanalyzed with the image subtraction
technique -- Difference Image Analysis ({\.Z}ebru{\'n}, Soszy{\'n}ski
and Wo{\'z}niak 2001a). Variable stars detected in that study were
presented in the catalog of variable stars in the Magellanic Clouds
({\.Z}ebru{\'n} \etal 2001b).

The main aim of this paper is to provide a catalog of eclipsing binary
stars in the SMC based on the DIA photometry. The catalog contains 1351
stars, from which only 896 were cross-identified in previous version
of the catalog indicating that 455 stars are newly discovered eclipsing
binary stars. 

The search algorithm and classification method were identical with those
used in the catalog of eclipsing binary stars in the LMC (Wyrzykowski
\etal 2003). We used artificial neural network for recognition of the
variability type and divided discovered eclipsing binaries into three
classical types: EA (Algol type), EB ($\beta$~Lyr type) and EW (W~UMa
type). The sample is reasonably complete at the level of the DIA catalog
of variable stars in the SMC although the completeness of the latter
drops rapidly for fainter objects. The sample allows statistical
analysis of eclipsing binaries in the SMC and should provide a good
material for testing theory of evolution of binary systems as well as
for studying the evolution of the SMC, star formation or other projects.
 
\Section{Observational Data}
All photometric data presented in the catalog of eclipsing stars were 
collected with the 1.3-m Warsaw telescope at the Las Campanas
Observatory, Chile, which is operated by the Carnegie Institution of
Washington, during the  second phase of the OGLE experiment. The
telescope was equipped with the  ``first generation'' camera with the
SITe ${2048\times2048}$ CCD detector working in driftscan mode. The
pixel size was 24${\mu}$m giving the scale of 0.417 arcsec/pixel. 

Observations of the SMC were performed in the ``slow'' reading mode of
the CCD detector with the gain 3.8e\(^{-}\)/ADU and readout noise of
about 5.4~e$^-$\!. Details of the instrumentation setup can be found in
Udalski, Kubiak and Szyma{\'n}ski (1997b). 

Regular observations of the SMC fields started on June 26, 1997 and covered
about 2.4 square degrees of central parts of the SMC. Reductions of the
photometric data collected up to the end of May 2000 were performed with
the Difference Image Analysis (DIA) package (Wo{\'z}niak 2000,
{\.Z}ebru{\'n}, Soszy{\'n}ski and Wo{\'z}niak 2001a) and variable stars
candidates were published in the catalog of variable stars in the
Magellanic Clouds ({\.Z}ebru{\'n} \etal 2001b).

\MakeTable{rrr}{11cm}{Equatorial coordinates of the SMC fields}{
\hline
\multicolumn{1}{c}{Field}&\multicolumn{1}{c}{RA (J2000)}&
\multicolumn{1}{c}{DEC (J2000)}\\
\hline
SMC\_SC1  & 0\uph37\upm51\ups & $-73\arcd29\arcm40\arcs$\\
SMC\_SC2  & 0\uph40\upm53\ups & $-73\arcd17\arcm30\arcs$\\
SMC\_SC3  & 0\uph43\upm58\ups & $-73\arcd12\arcm30\arcs$\\
SMC\_SC4  & 0\uph46\upm59\ups & $-73\arcd07\arcm30\arcs$\\
SMC\_SC5  & 0\uph50\upm01\ups & $-73\arcd08\arcm45\arcs$\\
SMC\_SC6  & 0\uph53\upm01\ups & $-72\arcd58\arcm40\arcs$\\
SMC\_SC7  & 0\uph56\upm00\ups & $-72\arcd53\arcm35\arcs$\\
SMC\_SC8  & 0\uph58\upm58\ups & $-72\arcd39\arcm30\arcs$\\
SMC\_SC9  & 1\uph01\upm55\ups & $-72\arcd32\arcm35\arcs$\\
SMC\_SC10 & 1\uph04\upm51\ups & $-72\arcd24\arcm45\arcs$\\
SMC\_SC11 & 1\uph07\upm45\ups & $-72\arcd39\arcm30\arcs$\\
\hline
}

The DIA photometry is based on the {\it I}-band observations. The
catalog of variable stars contains about 15\,000 stars in 11 fields of
the SMC (Table~1). Each star has at least 300 good photometric
measurements. The magnitudes of stars were transformed to the standard
system (Udalski \etal 2000). The errors of the measurements are about
0.005 mag for the brightest stars (${I<16}$~mag)  and grow to 0.08~mag
at 19~mag and to 0.3~mag at 20.5~mag.

\Section{Search and Classification of Eclipsing Binary Stars}
In order to identify eclipsing binaries we applied here the same
algorithm as used in the LMC search Wyrzykowski \etal (2003).
The main ideas employed are summarized below.

Automated recognition of the variability type is based on  artificial
neural network algorithm. Among all 15\,000 variable stars in the SMC we
selected only periodic ones, using the {\sc AoV} algorithm
(Schwarzenberg-Czerny 1989). Next, phased light curves of those stars
were transformed to ``images'', which were the network input.

Because our data come from the same source as in Wyrzykowski \etal
(2003), \ie the DIA catalog of variable stars ({\.Z}ebru{\'n} \etal
2001b) we did not have to repeat network learning process and used the
network with neural weights set previously for the LMC search. That
network was adopted to recognize the basic variability types, as
eclipsing, sinusoidal and ``saw shaped''. 

All light curves, which were classified by the network as eclipsing
ones, were inspected visually. Then the detected periods were tuned up
to smooth the eclipse shape which is very sensitive to period
inaccuracies. During visual inspection we divided eclipsing binary
stars into three classical types: EA (Algol type), EB ($\beta$~Lyr type)
and EW (W~UMa type), according to the Fourth Edition of ``General
Catalog of Variable Stars'' (GCVS, Kholopov \etal 1999). For several
stars multiple classification (\eg EB/EW) was chosen, because of
difficulties with distinguishing between those two classes. In the case
of several stars their variability, classification or period are
uncertain. Such objects are marked with additional remark as 
``uncertain''. 

Very uncertain objects were excluded at this stage. We also excluded
almost 100 objects, which were probably ellipsoidal variables. They were
automatically classified as eclipsing binary stars, because the shape of
their light curves revealed somewhat different depths of minima. However
in the case of some stars we were still unable to clearly distinguish
between eclipsing and ellipsoidal variables. Therefore, they are
additionally marked with ``ELL''. 

In the case of some objects additional variability of one or both
components was superimposed on the clear eclipsing variability. These
light variations could be caused by \eg spots on binary stars,
high proper motion of the system (long term falling or rising tendency
in the DIA light curve, see Soszy{\'n}ski \etal 2002), variability of
blending stars or probably by pulsations of one of the binary
components. All variables with  additional, confirmed or only suspected,
light curve changes are marked with  ``Puls'' or ``Puls?'' remark,
respectively. We did not perform any search for other periods in the
light curves of those stars, however we provide this information for two
examples. The first object is the star, noted also in Graczyk (2003),
with the ID number 661 in our catalog (SMC\_SC6 OGLE005139.70-731844.8),
where EA class star with period equal to 5.72593 days is probably
blended with another EA star with period of 2.617744 days. Light curve
of another star no.~778 (SMC\_SC6 OGLE005253.03-731111.8), exhibits also
two periodicities: 1.25171 days and 1.51228 days (EW and EA class
respectively). 

Additionally, we found 224 eclipsing variables with effects of orbit
eccentricity visible in their light curves. They were marked with
``ecc'' remark. In 49 cases we could not smooth both eclipses using the
same period what could be caused by large apsidal motion. We marked
these objects as ``eccAP'' and selected the period that smooths the
primary minimum. 

\renewcommand{\TableFont}{\tiny}
\renewcommand{\arraystretch}{1.35}
\MakeTableSep{r
@{\hspace{7pt}}r
@{\hspace{7pt}}r
@{\hspace{7pt}}r
@{\hspace{7pt}}r
@{\hspace{7pt}}r
@{\hspace{7pt}}r
@{\hspace{7pt}}r
@{\hspace{7pt}}r
@{\hspace{7pt}}r
@{\hspace{7pt}}r}
{11cm}{Eclipsing binaries in the SMC}{
\hline
\multicolumn{1}{c}{ No.}&
\multicolumn{1}{c}{Field}&
\multicolumn{1}{c}{Star}&
\multicolumn{1}{c}{OGLE}&
\multicolumn{1}{c}{~Period}&
\multicolumn{1}{c}{$T_0$}&
\multicolumn{1}{c}{$V$}&
\multicolumn{1}{c}{$B-V$}&
\multicolumn{1}{c}{$V-I$}&
\multicolumn{1}{c}{$I_{\rm PRI}$}&
\multicolumn{1}{c}{~~Type}\\
&&&
\multicolumn{1}{c}{No.}&
\multicolumn{1}{c}{~[days]}&
\multicolumn{1}{c}{--2450000}&
&
&
\multicolumn{2}{r}{(DIA)}
&\\
\noalign{\vskip5pt}
\hline
\noalign{\vskip5pt}
 1&SMC\_SC1&OGLE003617.48-731331.9&--    & 249.189650& 417.85070& 16.74&$-99.99$&$  1.42$& 0.20&EB\\
 2&SMC\_SC1&OGLE003618.10-733315.2&--    &   1.602530& 630.75054& 17.07&$-99.99$&$ -0.15$& 0.28&EA\\
 3&SMC\_SC1&OGLE003621.53-732610.6&12977 &  60.387530& 664.89765& 14.64&$  0.21$&$  0.27$& 0.54&EB\\
 4&SMC\_SC1&OGLE003639.62-730158.6&25589 &   3.396080& 625.77915& 17.29&$  0.01$&$  0.01$& 0.27&EA-ecc\\
 5&SMC\_SC1&OGLE003654.82-732625.7&13290 &   0.411280& 627.79314& 20.65&$-99.99$&$  2.92$& 0.42&EA\\
 6&SMC\_SC1&OGLE003655.53-734219.4&6130  &   1.914720& 625.93876& 17.64&$ -0.11$&$ -0.08$& 0.25&EA\\
 7&SMC\_SC1&OGLE003713.22-733244.3&--    &   1.253270& 628.54204& 17.02&$ -0.14$&$ -0.11$& 0.13&EB\\
 8&SMC\_SC1&OGLE003716.81-731602.1&--    &  15.712600& 629.93546& 16.73&$-99.99$&$  0.26$& 0.22&EB\\
 9&SMC\_SC1&OGLE003725.33-733016.1&--    & 145.403380& 585.68413& 18.53&$  0.84$&$  1.04$& 0.16&EA-uncertain\\
10&SMC\_SC1&OGLE003726.44-731547.3&--    &   3.116580& 628.96906& 18.57&$-99.99$&$  0.06$& 0.44&EA-eccAP\\
11&SMC\_SC1&OGLE003732.40-735619.2&27259 &   2.656750& 626.97921& 17.80&$ -0.08$&$  0.02$& 0.64&EB\\
12&SMC\_SC1&OGLE003736.43-733820.3&35641 &  29.056120& 645.32761& 18.50&$  0.37$&$  0.80$& 0.23&EB/ELL\\
13&SMC\_SC1&OGLE003738.86-734631.1&32309 &   3.819180& 624.82105& 16.35&$ -0.06$&$ -0.08$& 0.47&EB\\
14&SMC\_SC1&OGLE003744.93-734921.6&30716 &   2.449180& 626.51258& 17.36&$ -0.08$&$ -0.03$& 0.78&EB\\
15&SMC\_SC1&OGLE003753.54-732617.0&71827 &   1.646720& 621.34310& 18.45&$ -0.09$&$ -0.00$& 0.67&EA\\
16&SMC\_SC1&OGLE003759.97-732502.4&73863 &   6.215700& 616.99756& 19.62&$  0.61$&$  0.95$& 0.72&EA\\
17&SMC\_SC1&OGLE003804.71-735150.5&59297 &   3.796290& 623.29900& 19.41&$  0.04$&$  0.28$& 1.31&EA\\
18&SMC\_SC1&OGLE003805.03-731318.8&80268 &   1.521600& 622.09109& 17.79&$ -0.10$&$ -0.03$& 0.46&EA\\
19&SMC\_SC1&OGLE003814.59-730221.9&86988 &  70.299220& 723.86261& 18.31&$-99.99$&$  1.12$& 0.37&EA\\
20&SMC\_SC1&OGLE003831.81-733308.7&69238 &   1.452910& 620.89144& 17.02&$ -0.16$&$ -0.12$& 0.20&EA\\
21&SMC\_SC1&OGLE003835.24-735413.2&58930 &   0.269090& 621.77443& 15.89&$  0.97$&$  1.04$& 0.34&EW\\
22&SMC\_SC1&OGLE003836.68-732607.6&73575 &   1.400540& 620.97823& 18.30&$ -0.10$&$  0.00$& 0.46&EA\\
23&SMC\_SC1&OGLE003838.15-730953.8&--    &   3.589650& 619.51606& 19.48&$-99.99$&$  0.36$& 1.62&EA\\
24&SMC\_SC1&OGLE003843.72-732051.7&108663&   1.474950& 625.90916& 18.94&$-99.99$&$  0.20$& 0.95&EB\\
25&SMC\_SC1&OGLE003851.98-733433.2&99121 &   2.458900& 619.06728& 16.18&$ -0.16$&$ -0.10$& 1.05&EB\\
26&SMC\_SC1&OGLE003853.43-730323.8&--    &   0.554030& 621.13419& 18.49&$  0.07$&$  0.05$& 0.72&EA\\
27&SMC\_SC1&OGLE003858.13-732544.6&--    &   1.236260& 621.05691& 16.62&$ -0.05$&$  0.21$& 0.12&EB/EW/ELL\\
28&SMC\_SC1&OGLE003922.86-733905.2&--    & 102.126560& 598.11308& 17.33&$  1.38$&$  1.32$& 0.14&EB/ELL\\
29&SMC\_SC1&OGLE003924.70-732912.5&--    & 295.000000& 790.84736& 16.56&$  1.24$&$  1.35$& 0.06&EB\\
30&SMC\_SC1&OGLE003925.33-733835.9&97290 &   3.399940& 618.48291& 16.49&$ -0.13$&$ -0.10$& 0.26&EA-ecc\\
31&SMC\_SC2&OGLE003922.86-733905.2&--    & 102.126560& 597.49608& 17.35&$  1.40$&$  1.32$& 0.15&EB/ELL\\
32&SMC\_SC2&OGLE003925.33-733835.9&2742  &   3.399940& 628.68054& 16.49&$ -0.16$&$ -0.11$& 0.29&EA-ecc\\
33&SMC\_SC2&OGLE003927.37-733309.6&--    &  54.261380& 646.26467& 15.82&$-99.99$&$  0.09$& 0.21&EB/ELL+Puls\\
34&SMC\_SC2&OGLE003933.91-731855.4&11454 &   1.523980& 626.58340& 17.28&$-99.99$&$ -0.12$& 0.74&EA\\
35&SMC\_SC2&OGLE003937.81-732136.2&9700  &   0.886070& 626.78352& 17.82&$ -0.08$&$ -0.06$& 0.24&EW\\
36&SMC\_SC2&OGLE003941.37-725441.2&--    & 121.961620& 646.94259& 17.09&$-99.99$&$  1.40$& 0.08&EB/ELL\\
37&SMC\_SC2&OGLE003942.78-734330.1&102   &  23.919760& 644.67525& 18.31&$-99.99$&$  0.96$& 0.27&EB\\
38&SMC\_SC2&OGLE003946.11-733526.1&2788  &   1.977970& 625.55118& 16.38&$ -0.10$&$ -0.10$& 0.92&EB\\
39&SMC\_SC2&OGLE003952.15-730057.7&20205 &   1.006630& 626.53331& 18.77&$-99.99$&$  0.15$& 0.43&EW\\
40&SMC\_SC2&OGLE004000.10-733814.0&2917  &   1.308830& 628.95782& 18.21&$ -0.08$&$ -0.00$& 0.43&EA-ecc\\
41&SMC\_SC2&OGLE004002.96-732703.4&7640  &   0.857110& 626.07870& 17.20&$ -0.12$&$ -0.10$& 0.65&EW\\
42&SMC\_SC2&OGLE004003.31-733722.5&--    &   2.823900& 627.75032& 16.00&$ -0.11$&$ -0.13$& 0.12&EA/EB\\
43&SMC\_SC2&OGLE004009.01-733857.3&27167 &   2.368930& 626.10481& 16.45&$ -0.11$&$ -0.06$& 0.19&EB/EW/ELL\\
44&SMC\_SC2&OGLE004010.63-730120.1&--    &  50.316720& 583.38676& 17.65&$  0.26$&$  0.53$& 0.54&EA\\
45&SMC\_SC2&OGLE004024.03-732227.0&35961 &  11.896940& 619.67718& 19.86&$  0.89$&$  1.09$& 0.60&EA\\
46&SMC\_SC2&OGLE004028.41-733757.4&28572 &   1.842440& 620.84918& 16.95&$ -0.14$&$ -0.04$& 0.11&EB/ELL\\
47&SMC\_SC2&OGLE004037.20-732757.7&--    &   1.232960& 621.58291& 19.13&$  0.03$&$  0.18$& 1.57&EA\\
48&SMC\_SC2&OGLE004039.89-733409.1&29990 &   5.911090& 629.68464& 16.94&$  0.04$&$ -0.05$& 0.14&EA-ecc\\
49&SMC\_SC2&OGLE004045.39-730205.1&--    &  47.399640& 627.82925& 18.59&$-99.99$&$  0.93$& 0.49&EA\\
50&SMC\_SC2&OGLE004046.15-730833.0&--    &   1.968190& 621.62198& 19.03&$-99.99$&$  0.26$&1.52&EB\\
\hline}

\renewcommand{\arraystretch}{1.25}
\renewcommand{\TableFont}{\scriptsize}
\MakeTable{l@{\hspace{3pt}}rl@{\hspace{3pt}}r}{11cm}{Cross-identification 
of eclipsing binary stars detected in overlapping regions}{
SMC\_SC2 $\leftrightarrow$ SMC\_SC1 & OGLE003922.86-733905.2 &
SMC\_SC2 $\leftrightarrow$ SMC\_SC1 & OGLE003925.33-733835.9 \\
SMC\_SC2 $\leftrightarrow$ SMC\_SC3 & OGLE004225.74-732930.7 &
SMC\_SC3 $\leftrightarrow$ SMC\_SC4 & OGLE004524.54-732236.1 \\
SMC\_SC3 $\leftrightarrow$ SMC\_SC4 & OGLE004526.64-732531.1 &
SMC\_SC3 $\leftrightarrow$ SMC\_SC4 & OGLE004527.18-731549.8 \\
SMC\_SC3 $\leftrightarrow$ SMC\_SC4 & OGLE004528.58-730301.8 &
SMC\_SC3 $\leftrightarrow$ SMC\_SC4 & OGLE004528.81-730611.2 \\
SMC\_SC3 $\leftrightarrow$ SMC\_SC4 & OGLE004529.28-731006.9 &
SMC\_SC3 $\leftrightarrow$ SMC\_SC4 & OGLE004530.36-730331.9 \\
SMC\_SC4 $\leftrightarrow$ SMC\_SC3 & OGLE004533.56-730528.3 &
SMC\_SC4 $\leftrightarrow$ SMC\_SC3 & OGLE004534.08-731816.7 \\
SMC\_SC4 $\leftrightarrow$ SMC\_SC5 & OGLE004825.96-731745.8 &
SMC\_SC4 $\leftrightarrow$ SMC\_SC5 & OGLE004827.65-732141.3 \\
SMC\_SC4 $\leftrightarrow$ SMC\_SC5 & OGLE004828.65-731348.8 &
SMC\_SC4 $\leftrightarrow$ SMC\_SC5 & OGLE004828.90-731234.6 \\
SMC\_SC5 $\leftrightarrow$ SMC\_SC4 & OGLE004834.80-730652.6 &
SMC\_SC5 $\leftrightarrow$ SMC\_SC4 & OGLE004836.63-733531.3 \\
SMC\_SC5 $\leftrightarrow$ SMC\_SC6 & OGLE005125.57-731258.2 &
SMC\_SC5 $\leftrightarrow$ SMC\_SC6 & OGLE005126.84-731314.8 \\
SMC\_SC5 $\leftrightarrow$ SMC\_SC6 & OGLE005127.02-731307.4 &
SMC\_SC5 $\leftrightarrow$ SMC\_SC6 & OGLE005128.13-731517.6 \\
SMC\_SC5 $\leftrightarrow$ SMC\_SC6 & OGLE005129.62-732137.7 &
SMC\_SC5 $\leftrightarrow$ SMC\_SC6 & OGLE005130.22-725433.9 \\
SMC\_SC5 $\leftrightarrow$ SMC\_SC6 & OGLE005131.91-724538.7 &
SMC\_SC5 $\leftrightarrow$ SMC\_SC6 & OGLE005134.05-724626.1 \\
SMC\_SC5 $\leftrightarrow$ SMC\_SC6 & OGLE005134.85-724545.9 &
SMC\_SC6 $\leftrightarrow$ SMC\_SC5 & OGLE005135.04-731711.5 \\
SMC\_SC6 $\leftrightarrow$ SMC\_SC5 & OGLE005135.20-730420.8 &
SMC\_SC6 $\leftrightarrow$ SMC\_SC5 & OGLE005135.64-725432.0 \\
SMC\_SC6 $\leftrightarrow$ SMC\_SC5 & OGLE005135.81-731244.8 &
SMC\_SC6 $\leftrightarrow$ SMC\_SC5 & OGLE005136.23-732231.6 \\
SMC\_SC6 $\leftrightarrow$ SMC\_SC6 & OGLE005137.15-730550.2 &
SMC\_SC6 $\leftrightarrow$ SMC\_SC7 & OGLE005429.00-731846.2 \\
SMC\_SC6 $\leftrightarrow$ SMC\_SC7 & OGLE005431.85-723510.9 &
SMC\_SC6 $\leftrightarrow$ SMC\_SC7 & OGLE005432.05-725638.7 \\
SMC\_SC7 $\leftrightarrow$ SMC\_SC6 & OGLE005433.33-731315.0 &
SMC\_SC6 $\leftrightarrow$ SMC\_SC7 & OGLE005434.50-724051.9 \\
SMC\_SC7 $\leftrightarrow$ SMC\_SC6 & OGLE005434.68-725912.9 &
SMC\_SC7 $\leftrightarrow$ SMC\_SC6 & OGLE005437.09-730624.5 \\
SMC\_SC7 $\leftrightarrow$ SMC\_SC8 & OGLE005725.16-724738.7 &
SMC\_SC7 $\leftrightarrow$ SMC\_SC8 & OGLE005727.51-723514.7 \\
SMC\_SC8 $\leftrightarrow$ SMC\_SC7 & OGLE005734.96-725335.3 &
SMC\_SC8 $\leftrightarrow$ SMC\_SC9 & OGLE010023.48-725549.2 \\
SMC\_SC8 $\leftrightarrow$ SMC\_SC9 & OGLE010026.38-721503.4 &
SMC\_SC9 $\leftrightarrow$ SMC\_SC8 & OGLE010028.97-725918.5 \\
SMC\_SC9 $\leftrightarrow$ SMC\_SC10 & OGLE010318.41-723608.1 &
SMC\_SC9 $\leftrightarrow$ SMC\_SC10 & OGLE010319.89-722747.5 \\
SMC\_SC9 $\leftrightarrow$ SMC\_SC10 & OGLE010320.01-725004.3 &
SMC\_SC9 $\leftrightarrow$ SMC\_SC10 & OGLE010323.85-723010.9 \\
SMC\_SC10 $\leftrightarrow$ SMC\_SC11 & OGLE010616.65-724117.2 &
}

\Section{Catalog of Eclipsing Binary Stars}
In total 1351 eclipsing binary stars were found in the OGLE-II DIA
catalog of variable stars in the SMC fields. List of the first 50 stars
is presented in Table~2. It contains the ordinal number of the
eclipsing variable star, field, name of the star, number in the
previous version of the catalog, orbital period, heliocentric Julian
Date of the primary  minimum ($T_0$ -- 2\,450\,000, corrected for the
position of the star in the  driftscan image, as described in
{\.Z}ebru{\'n} \etal 2001b), {\it V}-band magnitude, ${B-V}$ and
${V-I}$ colors at maximum brightness from the standard OGLE-II data
pipeline PSF photometry, amplitude in the {\it I}-band from DIA 
photometry (depth of primary minimum) and eclipsing type. Color value of
${-99.99}$ indicates no observations in the {\it B} or {\it V} bands.

One should remember that the conversion of the DIA flux differences to
the magnitude scale is not always accurate. In particular, in the case
of severely  blended objects the depth of minima can be unreliable,
as the constant  flux cannot be accurately determined. Nevertheless,
such blends contain a real  eclipsing star. 

Among 1351 stars, 51 were identified twice in the overlapping regions
between the neighboring fields, therefore the total number of eclipsing
binary stars with independent measurements is equal to 1402. List of all
cross-identified objects is presented in Table~3. 

Because the DIA photometry data were taken from {\.Z}ebru{\'n} \etal
(2001b), stars' names follow their convention which is based on the
equatorial coordinates of the star for the epoch J2000 in the format:

\centerline{OGLE{\it hhmmss.ss-ddmmss.s}} 

\noindent For example, OGLE003617.48-731331.9 stands for a star with
coordinates ${\rm RA}=00\uph36\upm17\zdot\ups48$ and ${\rm
DEC=-73\arcd13\arcm31\zdot\arcs9}$. 

733 stars were classified as EA, 570 as EB and 150 as EW type. These
figures do not sum up to 1402, because of many multiple classifications.
Appendices A--C present examples of DIA {\it I}-band light curves of
types EA, EB and EW, respectively. The ordinate is the phase with 0.0
value corresponding to the deeper eclipse. Abscissa is the $I$-band
magnitude. Light curve is repeated twice for clarity.

Tables, light curves and finding charts of all 1402 eclipsing binary
objects are available from the OGLE {\sc Internet} archive and {\it via}
the WWW Interface (Section~7). 

Please note that periods of several stars might be two times longer than
the real one, because in the cases of faint stars and for noisy light
curves, the secondary eclipse could not be reliably detected. 

\Section{Discussion}
We present 1351 eclipsing binary stars located in the central regions of
the SMC found in the OGLE-II data collected during four observing
seasons. The number of stars is, however, smaller than found in the
previous release of the catalog (Udalski \etal 1998) based on the first
1.5 year of OGLE-II observations. This is likely due to incompleteness
of the DIA catalog of variable stars in the Magellanic Clouds
({\.Z}ebru{\'n} \etal 2001b), from which data for the present search
were taken.
\begin{figure}[htb] 
\vglue-2mm
\centerline{\includegraphics[width=10.3cm,height=10.3cm]{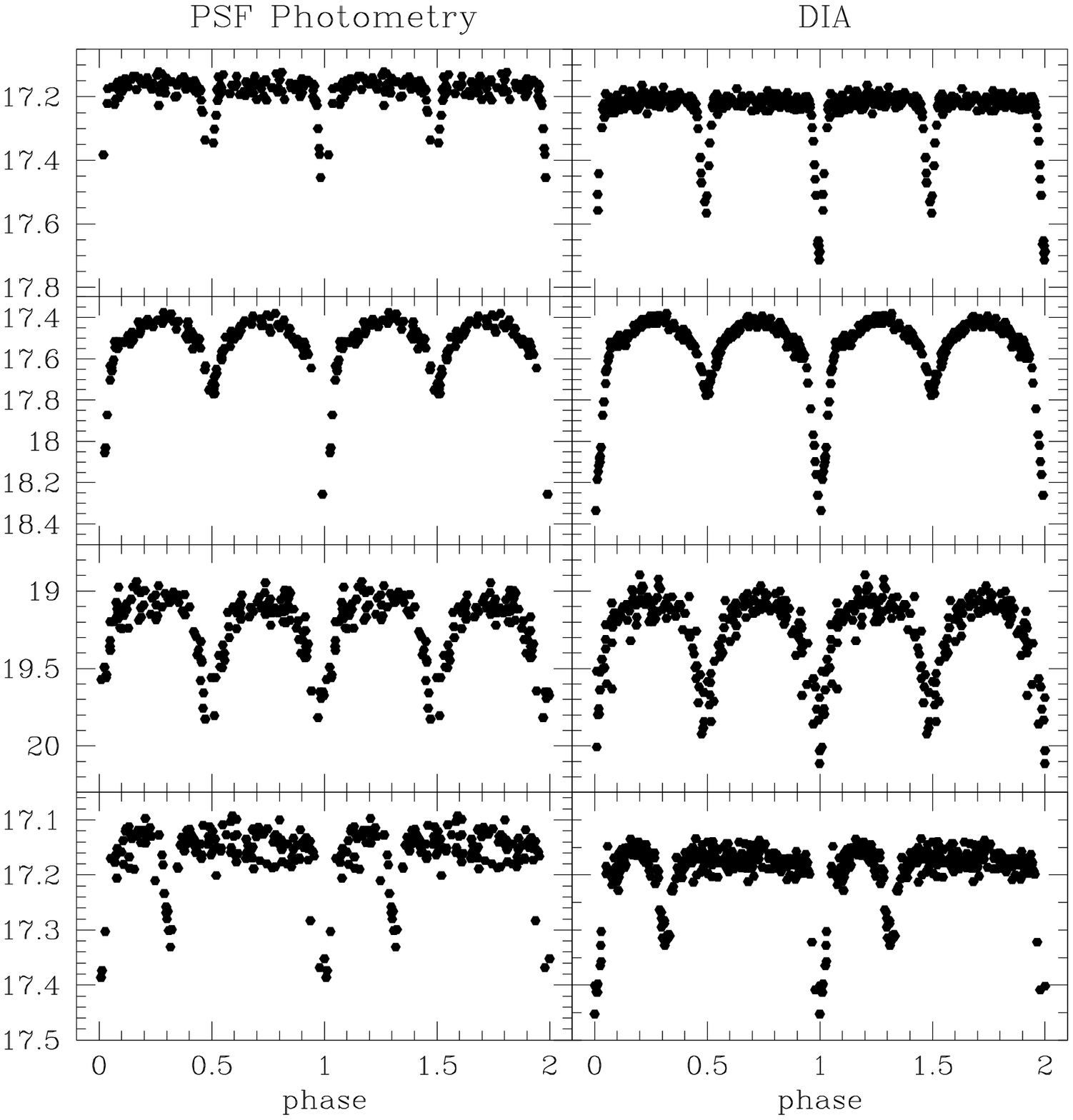}}
\vspace*{-3mm}
\FigCap{Comparison of light curves of the eclipsing binary stars from
the previous OGLE-II SMC catalog (Udalski \etal 1998) (left column) and
present version (right column).}
\end{figure} 

Only 896 stars were cross-identified with the previous catalog. It
means, that 455 stars presented here are newly discovered eclipsing
binary stars. The main improvements compared to the previous catalog
include much longer time-base of observations (4 years) and use of the
DIA photometry instead of convensional PSF photometry.  In the dense
stellar fields the former technique is superior to the latter. Overall,
present light curves have lower photometric scatter, better phase
coverage and they yield much more accurate periods.

A sample of light curves of eclipsing binary stars from both catalogs is
presented in Fig.~1. Stars from the catalog of Udalski \etal (1998) are
in the left column while the stars from the current catalog are in the
right column.  For some objects a small systematic shift up to about
0.05 mag  between light curves in the two catalogs can be seen. Most
likely it is due to differences in the zero point of calibrations or, as
mentioned above, inaccuracies in the conversion of the DIA flux
differences to the magnitude scale.

Fig.~2 presents histogram of the DIA $I$-band brightness for all eclipsing
binary stars found in the SMC (solid lines) and for those, which were
cross-identified with the previous edition of the catalog (dotted
line). The number of stars grows up to about ${I\approx18}$~mag and then
falls down to zero at ${I\approx20}$~mag. Newly discovered eclipsing binary
stars (difference of solid and dotted lines on the histogram) are
distributed more or less proportionally to the number of stars at a given
brightness.
\begin{figure}[p]
\centerline{\includegraphics[width=10cm, height=8cm]{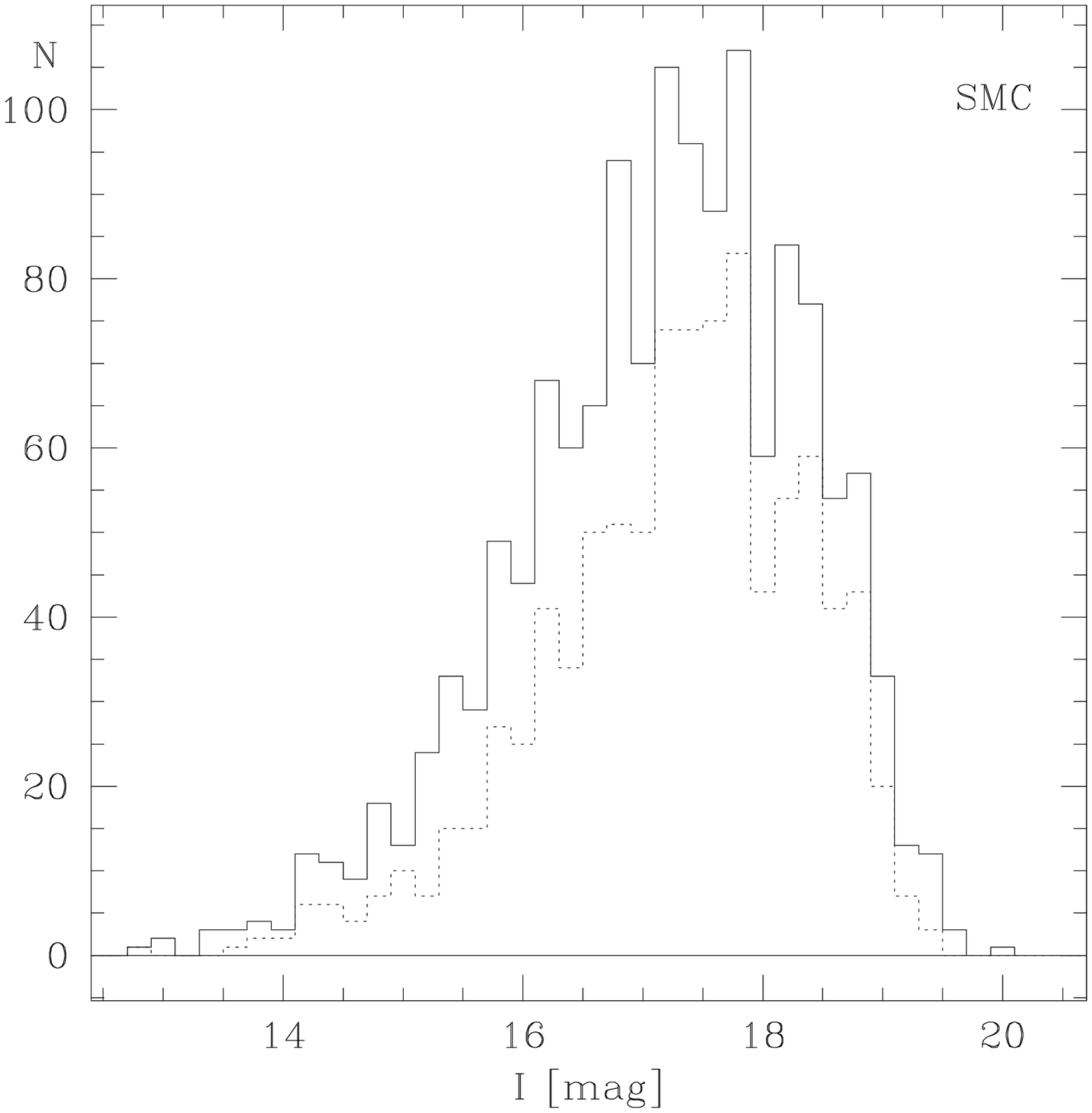}}
\FigCap{Histogram of the DIA $I$-band brightness in 0.2~mag bins for
all eclipsing binary stars (solid line) and only those, which were
cross-identified with the previous edition of the catalog (dotted line).}
\vskip5mm
\vskip6cm
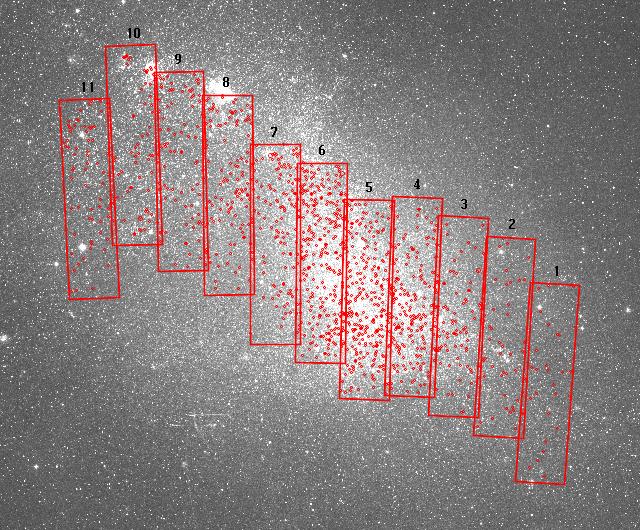
\vskip6cm
\FigCap{OGLE-II fields in the SMC. Dots indicate positions of eclipsing
stars. North is up and East to the left in the DSS image.}
\end{figure}
Fig.~3 presents a picture of the SMC from the Digitized Sky Survey (DSS)
with contours of the OGLE-II fields. Positions of the eclipsing binary
stars are marked with dots. The stars are distributed proportionally to
the density of the SMC stars, with the largest concentration in the
fields SMC\_SC4--SMC\_SC6. 

\begin{figure}[htb]
\centerline{\includegraphics[width=11cm,height=8.8cm]{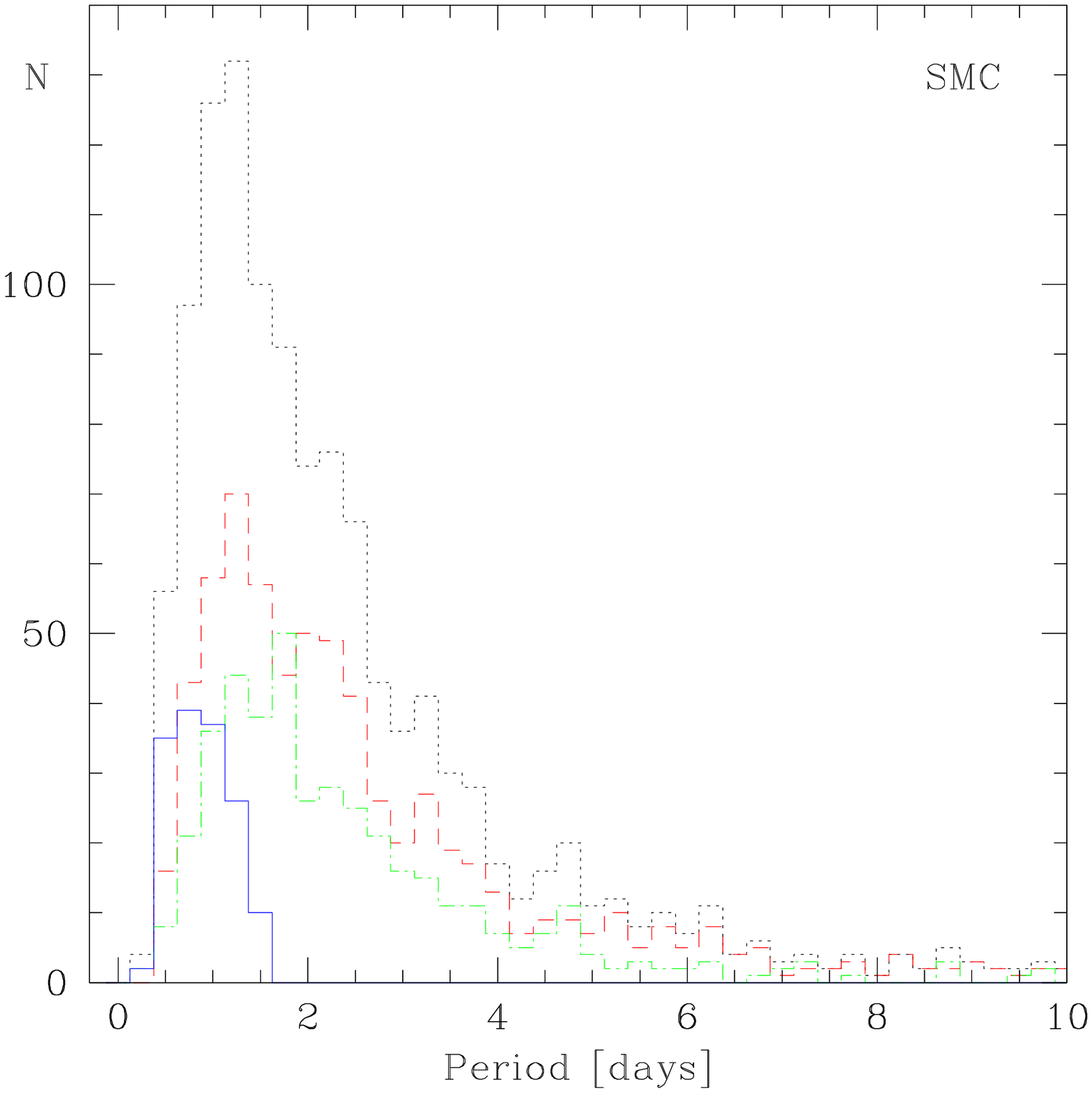}}
\FigCap{Histogram of periods of eclipsing binaries in 0.25~day bins.
Dashed (red), dot-dashed (green) and solid (blue) lines correspond to classes
EA, EB and EW respectively. Doted (black) line corresponds to all eclipsing
objects found in the SMC. Additional 235 objects have periods longer than
10 days.}
\end{figure} 

Fig.~4 shows the histogram of orbital periods of the SMC eclipsing stars
in 0.25~day bins from 0 to 10~days. Dashed (red), dot-dashed (green) and
solid (blue) lines
correspond to classes EA, EB and EW, respectively, and dotted (black) line
corresponds to all eclipsing objects. Additional 235 objects with
periods longer than 10 days are distributed more or less uniformly and
their number falls to zero at longer periods. The majority of stars are
short period systems with the most frequent period of about 1~day. The
longest period equals to 632.615 days (SMC\_SC3 OGLE004402.68-725422.5),
but both eclipses of this star are very similar and it is possible, that
the real period is twice that long if the star has very faint secondary
minima, invisible in our data. Another star (SMC\_SC4
OGLE004617.60-731859.0) has the period equal to 580.5 days.

\begin{figure}[htb]
\vglue-3mm
\centerline{\includegraphics[width=13cm]{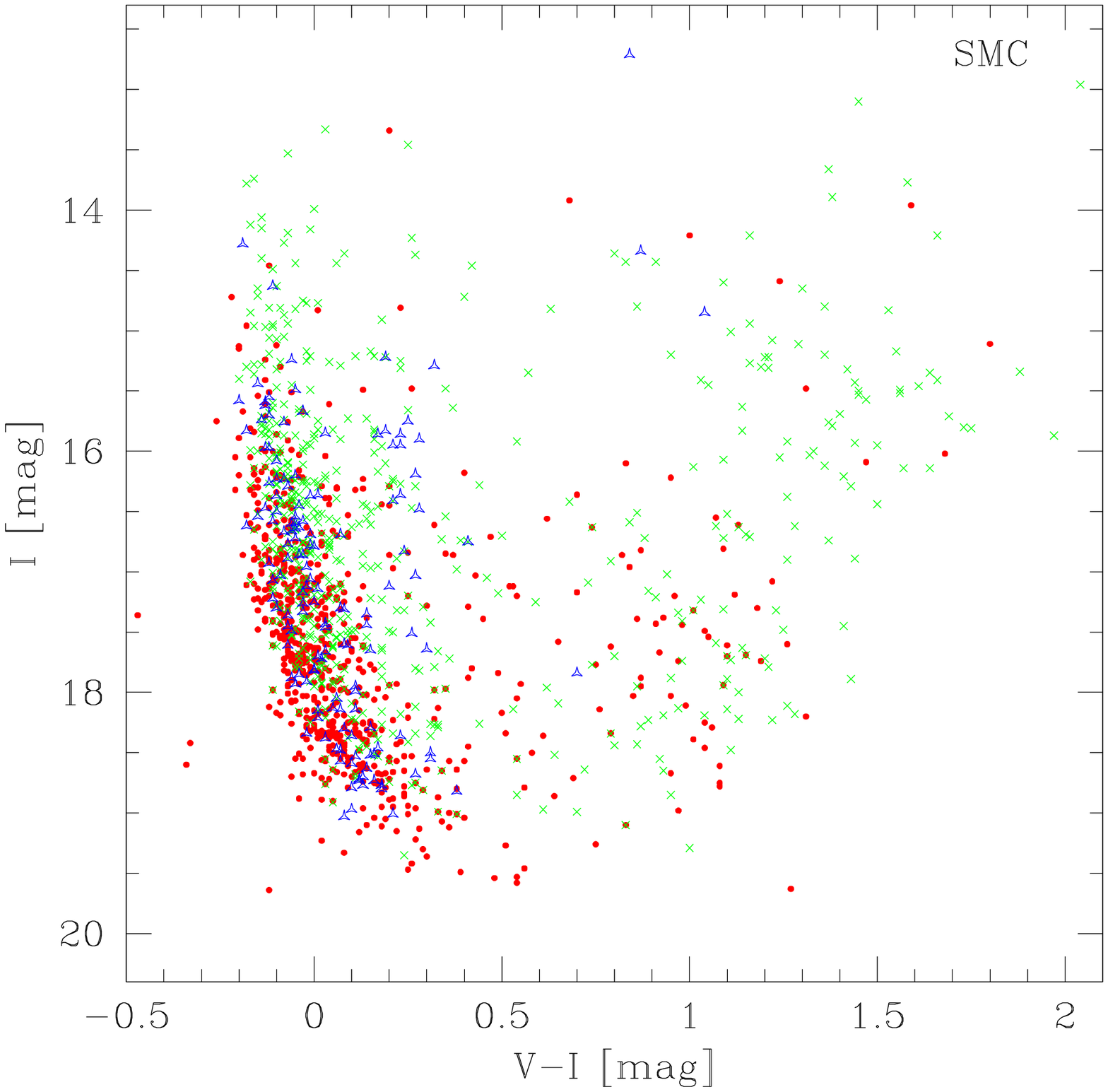}}
\FigCap{Color-magnitude diagram of eclipsing binaries in the SMC. Red dots,
 green crosses and blue triangles mark EA, EB and EW type objects,
 respectively.}
\end{figure} 
Fig.~5. presents {\it I} \vs ${V-I}$ color-magnitude diagram for all
eclipsing  binary stars from the catalog. EA, EB and EW classes are
marked with different symbols. Fig.~5 indicates, that most of the
eclipsing stars belong to the SMC, but there are also some foreground
stars, mostly EW class objects.

\begin{figure}[htb] 
\vglue-3mm
\centerline{\includegraphics[width=13cm]{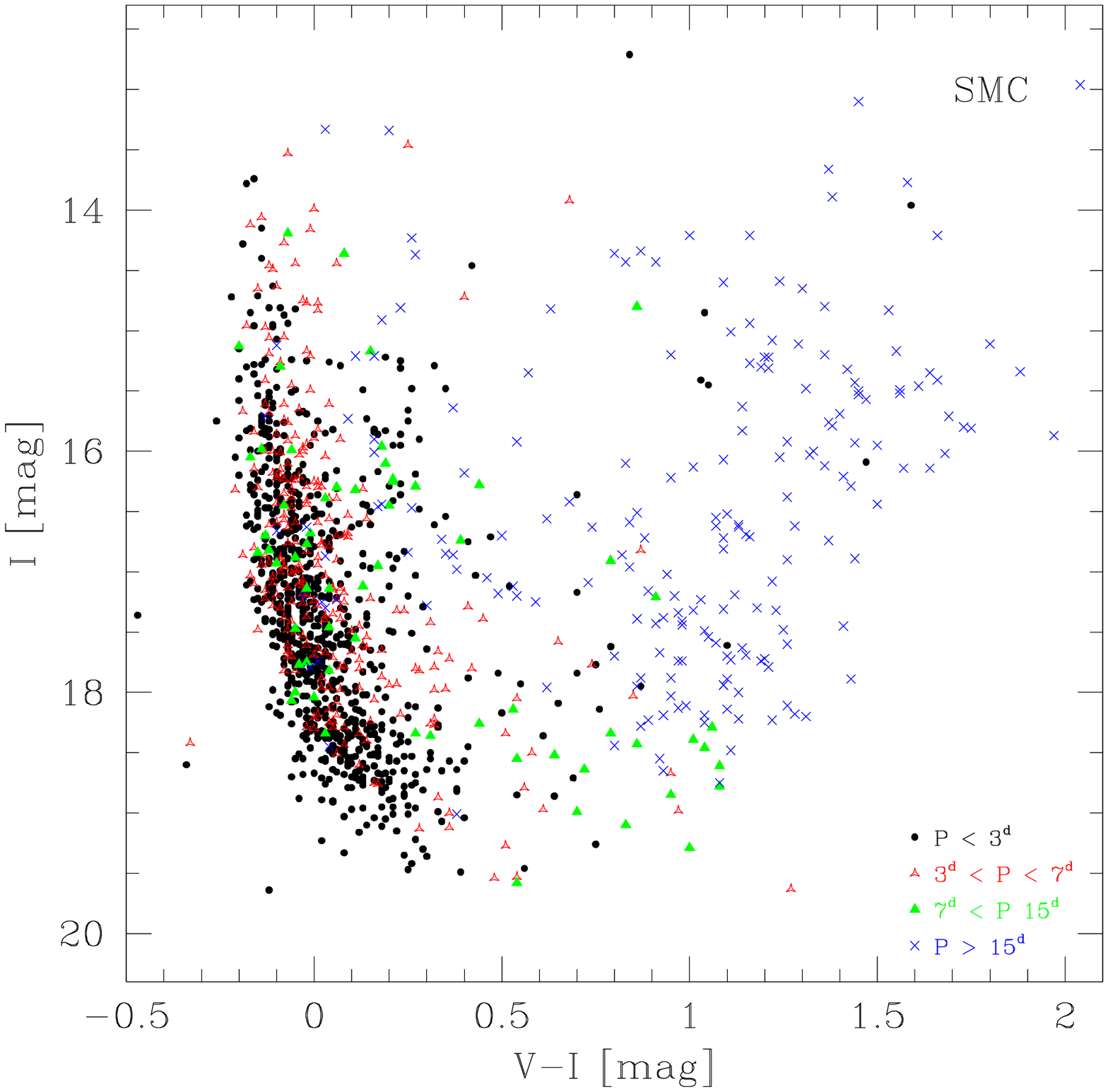}}
\FigCap{Color-magnitude diagram of eclipsing binaries in the SMC. Different 
symbols mark position of stars with short, medium, long and very long
periods.}
\end{figure} 
Another CMD diagram is presented in Fig.~6. Eclipsing binary stars are
divided into 4 groups depending on their periods: short (less than 3
days), medium (from 3 to 7 days), long (from 7 to 15 days) and very long
(more than 15 days). Each group is marked on the CMD with different
symbol and color. The majority of short and medium period eclipsing
stars are located on the main sequence and belong to the young
population of stars. Part of the long period stars are located also on
the main sequence, but some of them lie on the lower giant branch. Very
long period eclipsing stars are mostly concentrated on the red giant
branch.

\Section{Completeness of the Catalog}
We compared our catalog of eclipsing binary stars in the SMC with the
previous release by Udalski \etal 1998. They found 1527 objects (68
identified twice in overlapping fields) using the data covering first
1.5 year of observations of OGLE-II. Only 935 of them can be found in
the DIA catalog of variable stars ({\.Z}ebru{\'n} \etal 2001b), probably
because of incompleteness of the DIA catalog for faint stars. 30 out of
935 objects were not found in our preliminary results. In general, those
were stars with periods very close to 1, 2 or 5 days and were excluded
automatically before the network recognition process. Six objects had
very bad photometry or turned out not to be an eclipsing object.
Therefore, 24 stars  total were added to our final catalog. This yields
97\% efficiency of our search algorithm.

To estimate the completeness of our catalog we compared objects detected
in the overlapping regions of neighboring fields. Based on astrometric
solutions, we checked, which of the detected eclipsing binary stars
should have a counterpart in the neighboring field and compared these
objects with actually detected stars. In total, 102 stars should have a
``twin'' in the neighboring field. In practice 92 stars with pairs were
found, yielding the mean completeness of our catalog equal to about
90\%. This is certainly a lower limit value, because the edges of each
field can be affected by non-perfect pointing of the telescope leading to
effectively smaller number of observations and consequently to a smaller
probability of variability detection. 

Five missing pairs were subsequently added manually to the catalog. 
Parameters of all 102 paired stars were very similar to their ``twins'',
however we unified them to the values of the star with larger number of
data points. 

\Section{The Catalog in the I{\small NTERNET}}
The catalog of eclipsing binary stars is available on-line through {\sc
ftp} and WWW from the OGLE {\sc Internet} archive. The catalog can be
accessed {\it via} anonymous {\sc ftp} at the following addresses: 
\vskip3pt
\centerline{\it ftp://sirius.astrouw.edu.pl/ogle/ogle2/var\_stars/smc/ecl}
\centerline{\it ftp://bulge.princeton.edu/ogle/ogle2/var\_stars/smc/ecl}
\vskip3pt
\noindent
WWW interface to the catalog is available from the following addresses:
\vskip3pt
\centerline{\it http://ogle.astrouw.edu.pl}
\centerline{\it http://bulge.princeton.edu/\~{}ogle}

The catalog will be regularly updated when the final set of the OGLE-II
data becomes available and/or any errors (to some extent unavoidable in
a dataset so large) are found. The most recent version will be
available {\it via} {\sc Internet} from the above addresses. The catalog
will also be significantly extended when the number of epochs
in the ongoing OGLE-III phase becomes large enough to facilitate good
detection efficiency for eclipsing binaries. As the OGLE-III fields cover
practically entire SMC our goal is to conduct a complete census of
eclipsing stars in the SMC. 

\Section{Summary}
The new version of the catalog of eclipsing binary stars in the SMC
based on the OGLE-II DIA catalog of variable stars in the Magellanic
Clouds contains 1351 objects of three classical types: EA, EB and EW.
455 stars are newly discovered eclipsing binaries, not found in the
previous edition of the catalog (Udalski \etal 1998).  The exceptional
good quality of the DIA photometry and  very long time-base of the
OGLE-II observations  enabled construction of a uniform sample of
eclipsing binaries  with high quality light curves and very accurate
periods. The catalog provides observational material  for a variety of
astrophysical studies in the SMC.

\Acknow{We would like to thank Prof.~Bohdan Paczy{\'n}ski for his 
encouragements and discussions about this work. This work was partly
supported by the KBN grant 2P03D02523 to {\L}.~Wyrzykowski, 2P03D02124
to A. Udalski and NASA grant NAG5-12212 and NSF grant AST-0204908 to
B.~Paczy\'nski. We acknowledge usage of the Digitized Sky Survey which
was produced at the Space Telescope Science Institute based on
photographic data obtained using the UK Schmidt Telescope, operated by
the Royal Observatory Edinburgh.}

\begin{figure}[p]
\centerline{\large\bf Appendix A}
\vskip9pt
\centerline{\bf Eclipsing stars in the SMC}
\vskip5pt
\centerline{\bf EA type eclipsing stars}
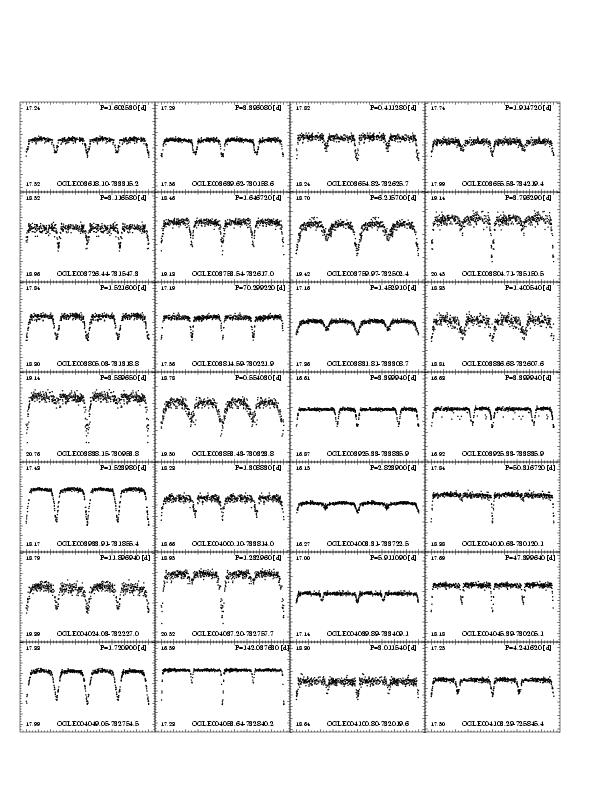
\end{figure} 
\begin{figure}[p]
\centerline{\large\bf Appendix B}
\vskip9pt
\centerline{\bf Eclipsing stars in the SMC}
\vskip5pt
\centerline{\bf EB type eclipsing stars}
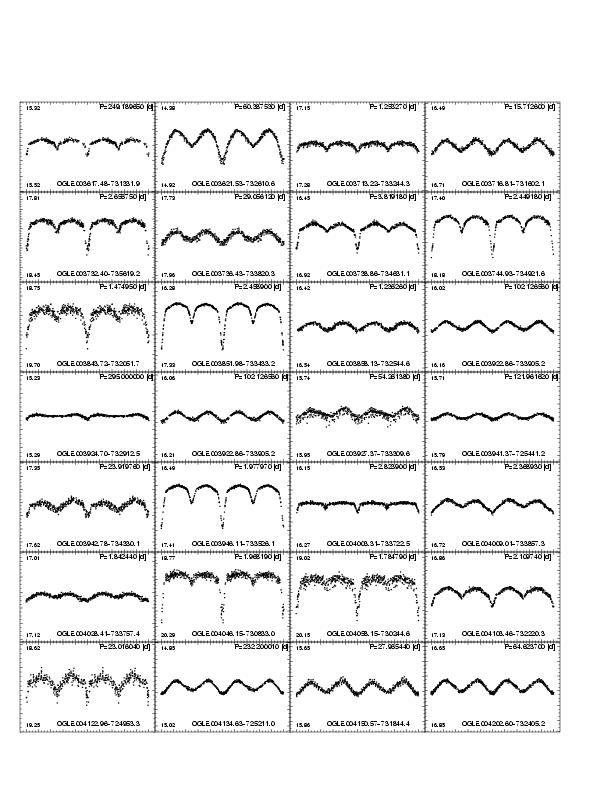
\end{figure} 
\begin{figure}[p]
\centerline{\large\bf Appendix C}
\vskip9pt
\centerline{\bf Eclipsing stars in the SMC}
\vskip5pt
\centerline{\bf EW type eclipsing stars}
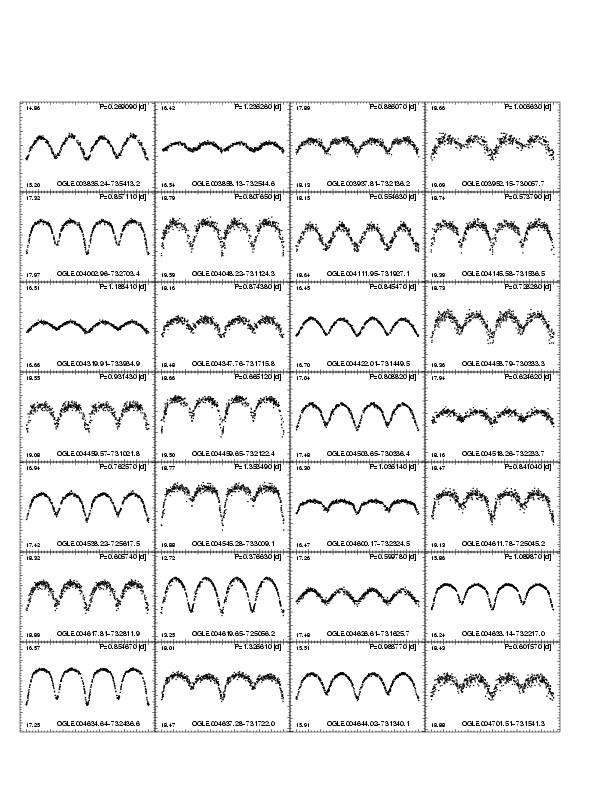
\end{figure} 
\end{document}